\documentstyle[12pt]{article}
\title{Some consequences of the effective low-energy lagrangian for gravity}
\author{{\large \bf A. Dobado and A. L\'opez} \\
Departamento de F\'{\i}sica Te\'orica \\
Universidad Complutense de Madrid\\
 28040 Madrid, Spain }
\begin{document}
\maketitle
\begin{abstract}

We consider the minimal low-energy action for gravity up to six derivatives
which is renormalizable at the two-loop level modulo higher
derivative corrections. Then we study the classical solutions corresponding
to the Schwarzschild and the Robertson-Walker metrics. In the first case we
find a singularity close to the gravitational radius and in the second case we
find inflationary  de Sitter solutions in absence of any matter or cosmological
constant.

 \end{abstract}
\vskip 3.0cm
FT/UCM/7/93
\newpage
%\baselineskip 0.83 true cm
%\textheight 20 true cm
\newpage
{\bf Introduction}

Today most of the physicist agree that the issue of quantum
gravity  is probably the
most important open problem in theoretical physics
(for a recent review see [1] and references therein). At the present moment it
is not clear if the formulation of the quantum theory of gravitation will
require going beyond quantum field theory like it happens in string theory or
even a deeper modification of quantum mechanics.

 Of
course this question will not be solved here, but until it is clarified
one can adopt a much more modest attitude trying to figure up which could
be the low-energy impact of the real theory of gravitation. It is clear that at
low energies this theory will lead to the standard Einstein lagrangian.
However, this lagrangian will in general be modified by higher-order terms in
the sense of the number of derivatives of the space-time metric over the
Planck mass scale $M_P$. Thus, the Einstein lagrangian with two derivatives
and one factor $M^2_P$ would be the first term, then we would have terms
with four derivatives and one $M_P^0=1$ factor, next we would have six
derivative terms with one $M^{-2}_P$ factor and so on. The form of all these
terms is dictated just by the number of the derivatives and general covariance.
Different underlying gravitational theories will predict different values for
the adimensional constants affecting these terms, being superstring theories a
typical example [2].

 Independently of the concrete value
of the constants in this low-energy effective lagrangian for gravitation, it is
clear that one can use this derivative expansion up to some number of
$\partial ^2/ M_P^2$ powers to predict the qualitative behavior of the
modifications to the standard Einstein gravitation theory. However, in order to
be completely consistent one should also take into account the effect of the
quantum corrections to this effective lagrangian. For example the loops
obtained with the Einstein lagrangian produce corrections which are of the
order of
 $(\partial ^2/ M_P^2)^2$, two-loop diagrams
give rise to corrections which are of the order of $(\partial ^2/ M_P^2)^3$
etc.

This framework for the effective description of the low-energy behavior of
gravity has some similitude with  that proposed by Weinberg some time ago
[3] for the description of the low-energy pion scattering. In that case
the effective lagrangian is also the most general lagrangian compatible with
the
 (approximate) chiral symmetry of the strong interactions $SU(2)_L\times
SU(2)_R$ which is supposed to be spontaneously broken to the group
$SU(2)_{L+R}$. Pions are understood as the Goldstone bosons associated to this
symmetry breaking and this  is why they are (nearly) massless. The
corresponding
fields $\pi_a$ are just coordinates parametrizing the coset space
$SU(2)_L\times
SU(2)_R/SU(2)_{L+R}$. This space  is   a three
sphere i.e. a maximally symmetric space having the chiral group $SU(2)_L\times
SU(2)_R$ as isometry group. At low energies the pion dynamics is controlled by
the effective lagrangian

\begin{equation}
{\cal L}_{eff}=\frac{1}{2}g_{ab}(\pi)\partial_{\mu}\pi^a
\partial ^{\mu}\pi^b+\alpha(g_{ab}(\pi)\partial_{\mu}\pi^a
\partial ^{\mu}\pi^b)^2+\beta(g_{ab}(\pi)\partial_{\mu}\pi^a
\partial ^{\nu}\pi^b)^2+ O(\partial ^6)
\end{equation}
where $g_{ab}$ is the metric on the
 sphere. Here again the quantum corrections to the two derivative term are
of the order of $O(\partial ^4)$ etc. In principle the constants $\alpha,
\beta...$ can be obtained from an underlying theory of the strong interactions
like QCD. Nevertheless, even in the case that this theory were not known or
the computation of the constants were hard or impossible, the above
lagrangian can be very usefully used by obtaining the constants from the
experiment and using them to make new predictions. This program was carried out
by Gasser and Leutwyler in a remarkable set of papers [4] to the one loop level
and ever since this kind of approach to the low-energy hadron interactions has
been followed by many particle physicists with great phenomelogical success.
More recently, a similar approach has been also applied to the phenomenological
description of the unknown symmetry breaking sector of the standard model and
in
particular to the parametrization of the elastic scattering of the longitudinal
components of the weak bosons at the $TeV$ scale [5].

          Following this very general line of thinking one could try to make an
effective low-energy description of gravity. However, some differences appear
when we compare pions or longitudinal components of the weak bosons with
gravitons. First we do not have any coset space to make a Goldstone boson like
interpretation of gravitons (see [6] for some attempt in this direction).
Second, gravity is a gauge theory and we need to go through the standard
procedure of quantization of these theories including gauge fixing and ghost
terms in the lagrangian to relate properly the ten independent components
of the metric tensor with the two physical degrees of freedom of the graviton.
In spite of these facts, some analogies still remain with the case of the
non-linear sigma models. For example, as it was mentioned above, the one-loop
radiative corrections to the Einstein lagrangian  have four derivatives.
The divergent terms of these corrections were computed some time ago by
't Hooft and Veltman [7]. Using the background field method and
dimensional regularization  it is possible to find:
 \begin{equation}
{\cal L}_{eff}^{(1)}=\frac{1}{(4\pi) ^2\epsilon}(\frac{R^2}{120}+\frac{7}{20}
R^{\alpha\beta}R_{\alpha\beta})
\end{equation}
 The important point here is that these divergences do not contribute to
the $S$ matrix elements (note that they vanish on shell)  so that, the
Einstein theory of gravitation considered as a standard quantum field
theory is one-loop finite. In fact there is a simple argument to see that
it must be so.  The  four-derivative local terms that in principle can be
generated by the one loop computation are $R^2$,
 $R^{\alpha\beta}R_{\alpha\beta}$ and
$R^{\alpha\beta}_{\;\;\;\;\gamma\delta}R^{\gamma\delta}_{\;\;\;\;\alpha\beta}$.
Only the last one does not vanish on shell, but as it
is well known the combination
$-4R^{\alpha\beta}_{\;\;\;\;\gamma\delta}R^{\gamma\delta}_{\;\;\;\;\alpha\beta}
+16R^{\alpha\beta}R_{\alpha\beta}-4R^2$ is a total derivative in four
dimensions related (for Euclidean signature) with the Euler class of the
space-time manifold. Thus, modulo a renormalization of this topological term,
the one-loop divergences of pure Einstein gravity vanish on shell.

More recently, the two-loop  divergent contribution to the effective action
for gravity was computed by Goroff and Sagnotti [8]. In this huge
computation they were forced to use an special $C$ code since standard
algebraic
manipulators  were unable to work it out completely (see [9] for a previous
disscussion about this point). The on-shell contribution to the effective
lagrangian coming from the Goroff and Sagnotti computation is given by:

\begin{equation}
{\cal
L}_{eff}^{(2)}=\frac{209}{360(4\pi)^3M_P^2\epsilon}R^{\alpha\beta}_{\;\;\;\;\gamma\delta}R^{\gamma\delta}_{\;\;\;\;\sigma\rho}
R^{\sigma\rho}_{\;\;\;\;\alpha\beta} \end{equation}
Therefore, this remarkable computation showed that the Einstein theory of
gravitation is not renormalizable in the common sense. However, one can still
consider it as the first term of a derivative expansion by following the
philosophy of the phenomenological lagrangians. The new  terms that must be
added to some given order in the number of derivatives play a double role.
First, they have coefficients that carry the information about the underlying
theory of gravitation and second, they are also counter terms that absorb the
divergences. It could be possible in principle to compute the low-energy
action of gravitation in this way up to some number of derivatives i.e. powers
of the external momenta, in terms of some renormalized constants that in
principle could be obtained from the experiments or observations.. Once they
were fitted, this effective action could be used to make new predictions

Obviously, this approach has many difficult problems that must be solved.
First,
from the theoretical point of view, the practical computation is very hard.
Note that we only know the divergent part of the one and two-loop effective
action  but we even do not know the one-loop finite part contribution.
This kind of contribution is in general non-local since gravitons are
massless. Second, we have of course phenomenological problems. As the higher
derivative terms are suppressed by powers of  $M_P$ they become practically
unobservable. The only possibility for these terms to be relevant is in
physical contexts where the curvature is very large. Typically these
situations occur in the regions close to the singularities appearing in the
classical solutions of the  standard theory and in fact it could be the case
that the effective low-energy theory of gravitation  completely avoids these
singularities solving one of the old problems of the Einstein theory (we
will see later that this is not necessarily the case).

Motivated by this interesting possibility we have started the program of
studying the effect on the classical solutions of the gravitation theory
including the higher local derivative terms that appear in the low-energy
effective action for gravity. In fact some of these terms have already been
considered in the literature, typically the four derivative ones.
However, here we will adopt a different attitude. As the general analysis is in
principle very complex we want to select the minimal set of terms that must be
added to the standard lagrangian to have  a sensible low-energy effective
lagrangian. From the Goroff and Sagnotti work we know that, at the
 six derivatives level, the only term we need to add to get
an on-shell finite effective action is that of eq.3. Therefore, we can
conclude that the addition of this term to the standard lagrangian produces the
minimal consistent effective low-energy two-loop renomalizable lagrangian  in
the sense of the effective  theories considered in  [3] i.e. neglecting higher
derivative terms. For the sake of simplicity this will be the only one
considered in this note, so that we pass by the more commonly considered in the
literature four derivative terms. In this note we will explore the effect of
the six derivative term on the important Schwarzschild and Robertson-Walker
metrics.
\vskip 1.0cm

{\newpage}

{\bf The
equations of motion}

According to the above discussion we will consider the effective action:

\begin{equation}
\Gamma_{eff}[g_{\alpha\beta}]=\int d^4x\sqrt{g}{\cal L }_{eff}
\end{equation}
where the effective lagrangian is given by:
\begin{equation}
{\cal L}_{eff}=-\frac{M_P^2}{16\pi}R+\frac{\alpha}{M_P^2}
R^{\alpha\beta}_{\;\;\;\;\gamma\delta}R^{\gamma\delta}_{\;\;\;\;\sigma\rho}
R^{\sigma\rho}_{\;\;\;\;\alpha\beta}
\end{equation}

where $R^{\alpha\beta}_{\;\;\;\;\gamma\delta}$ is the Riemann tensor,
$R$ is the
curvature scalar (we will follow the notation convention of the Weinberg's
book [10]), $M_P$ is the Planck mass and $\alpha$ is an adimensional coupling
constant. By applying the standard methods we can obtain the corresponding
equation of motion:
\begin{eqnarray}
\frac{M_P^2}{16\pi}(R^{\mu\nu}-\frac{1}{2}g^{\mu\nu}R)=
-\frac{\alpha}{M_P^2}(\frac{1}{2}g^{\mu\nu}
R^{\alpha\beta}_{\;\;\;\;\gamma\delta}R^{\gamma\delta}_{\;\;\;\;\sigma\rho}
R^{\sigma\rho}_{\;\;\;\;\alpha\beta} \end{eqnarray}
\begin{eqnarray}
\nonumber
-3R^{\alpha\mu}_{\;\;\;\;\gamma\delta}R^{\gamma\delta}_{\;\;\;\;\sigma\rho}
R^{\sigma\rho\;\;\nu}_{\;\;\;\;\alpha}+6(R^{\nu\delta}_{\;\;\;\;\sigma\rho}R^
{\sigma\rho\mu\beta})_{;\delta;\beta})\end{eqnarray}

When matter is present, we have to include a new piece in the effective
lagrangian ${\cal L}_m$ and as a consequence of that we must add a new
term in the RHS of the above equation of motion which is $-T^{\mu\nu}/2$ being
$T^{\mu\nu}$ the matter energy-momentum tensor. Obviously the Einstein
equations of motion are obtained just by
taking $\alpha=0$, but for general $\alpha$, the modified equations will
produce
new effects that will be studied in next section.
\vskip 1.0cm

 {\bf The modified Schwarzschild metric:}

The most general form of the static and isotropic metric can be written as:
\begin{equation}
d\tau^2=B(r)dt^2-A(r)dr^2-r^2(d\theta^2+sin^2\theta d\phi^2)
\end{equation}
	By inserting this metric in the eq.6, we
can find the equations to be satisfied by the $A(r)$ and $B(r)$ functions. For
the case $\alpha=0$, these equations were solved by
Schwarzschild in 1916 for the case of a flat  spatial infinity:
\begin{eqnarray} B(r)=1-\frac{2M}{rM_P^2} \\ \nonumber A(r)=1/B(r)
\end{eqnarray} being $M$ the total mass of the body producing the gravitational
field. For general $\alpha$, the task of finding the corresponding solutions is
very hard. However, following the phylosophy of the effective lagrangians, one
could try to find at least the first coefficients of the large distance
expansion of the $A(r)$ and $B(r)$ functions: \begin{eqnarray}
A(r)=\sum_{n=0}^{\infty}a_n(\frac{L_p}{r})^n \\   \nonumber
B(r)=\sum_{n=0}^{\infty}b_n(\frac{L_p}{r})^n
\end{eqnarray}
where $L_p$ is the Planck lenght $(L_p=1/M_P)$ and $a_n$ and $b_n$ are
adimensional coefficients with $a_0=b_0=1$. Introducing these expansions into
the equations of motion, we find that the first correction to the
Schwarzschild solution appears to be of the order of $(L_p/r)^6$. To find
these corrections, we can introduce the Schwarzschild solutions in the RHS
of the equation of motion and a completely general static, isotropic metric in
the LHS. With the help of a REDUCE code we find the following equations:
\begin{eqnarray}
-\frac{M_P^2}{16\pi}(\frac{1}{r^2AB}-\frac{A'}{rA^2B}-\frac{1}{r^2B})  =
\end{eqnarray}
\begin{eqnarray}
\nonumber
 \frac{\alpha}{M_P^2}(1-\frac{2M}{rM_P^2})^{-1}(\frac{2352(M/M_P^2)^3}{r^9}-
\frac{1080(M/M_P^2)^2}{r^8}) \end{eqnarray}
 \begin{eqnarray}
-\frac{M_P^2}{16\pi}(\frac{-B'}{rA^2B}-\frac{1}{r^2A^2}+\frac{1}{r^2A})  =
 \end{eqnarray}
\begin{eqnarray}
\nonumber
\frac{\alpha}{M_P^2}(1-\frac{2M}{rM_P^2})(\frac{240(M/M_P^2)^3}{r^9}-
\frac{216(M/M_P^2)^2}{r^8}) \end{eqnarray}
\begin{eqnarray}
-\frac{M_P^2}{16\pi}(\frac{A'}{2r^3A^2}-\frac{B'}{2r^3AB}-\frac{B''}{2r^2AB}
+\frac{(B')^2}{4r^2AB^2}+\frac{A'B'}{4r^2A^2B})  =\end{eqnarray}
\begin{eqnarray}
\nonumber
\frac{\alpha}{r^2M_P^2}(\frac{-1488(M/M_P^2)^3}{r^9}+
\frac{648(M/M_P^2)^2}{r^8})
\end{eqnarray}
where eq.10 corresponds to the temporal coordinate, eq.11 to the radial one and
eq.12 to the angular coordinates (both produce the same equation). Now, from
eq.10 we find the coefficients of the $A$ expansion, from eq.11 we find the $B$
coefficients and finally, eq.12 is found to be compatible with the other two.
The values  found for the coefficients go as follows:$a_0=1, a_1=2M/M_P,
a_k=a_1^k$ for $k=2,3,4,5$ and $a_k=a_1^k+16\pi\alpha a_1^{k-4}(54+(k-6)5)$ for
$k=6,7,8,9$ and $b_0=1, b_1=-a_1, b_k=0$ for $k=2,3,4,5,6, b_7=-16\pi\alpha
5a_1^3$ and $b_8=b_9=0$. As expected, the first corrections to the
Schwarzschild metric are of the order of $(L_p/r)^6$. Note also that now $A$
is not equal to $B^{-1}$ and this fact will have important consequences that
will be discussed later. With regard to the validity of the coefficients
obtained above, the approximation we have used introducing the Schwarzschild
metric in the RHS of eq.6 gives correct values for the coefficients up to
terms of order 9. However, when one consider the effect of possible
counterterms in the lagrangian with eight or more derivatives, they could
affect coefficients of terms of order seven or higher.

To study the effect of the new $\alpha$ term on the static isotropic
solution one could consider an alternative approach by making a
perturbative expansion in the $\alpha$ coupling constant instead of an
expansion in $L_p/r$. To the lowest order, it is possible to find:
\begin{equation}
A(r)=\frac{1}{1-\frac{2M}{rM_P^2}}+16\pi\alpha(\frac{2M}{M_P})^2
(\frac{1}{rM_P})^6(\frac{54}{1-\frac{2M}{rM_P^2}}+
\frac{10M/(rM_P^2)}{(1-\frac{2M}{rM_P^2})^2}) \end{equation} \begin{equation}
B(r)=1-\frac{2M}{rM_P^2}-80\pi\alpha (\frac{2M}{M_P})^3(\frac{1}{rM_P})^7
\end{equation}
 It is easy to see that the behaviour of this solution close to
the gravitational radius $r_s=2M/M_P^2$ is now controlled by the $\alpha$ term.
  For instance, it is now possible to find a value $r_A$ near $r_s$
where $A(r_A)=0$ but for  the ordinary Schwarzschild solution such a
point does not exist. In this case the $A$ function diverges   and the $B$
function vanishes at the
gravitational radius of the body i.e. at the event
horizon
 $A(r_s)=\infty$ and $B(r_s)=0$. Similarly, one could ask where this happens in
our modified solution. Then it is not difficult to see that for general
$\alpha$
the point $r_B$ where $B(r_B)=0$ is different from the point where $A$
diverges which continues being the gravitational radius $A(r_s)=\infty$. For
example, for $79\mid \alpha \mid \pi (M_P/M)^4 <<1$ (which is a very good
approximation for macroscopic bodies) we find, for positive $\alpha$, that
$r_B>r_s>r_A$ with $r_A \simeq(2M/M_P^2))(1 -\alpha 5 \pi(M_P/M)^4)$ and   $r_B
\simeq(2M/M_P^2))(1 +\alpha 5 \pi(M_P/M)^4)$ (where A vanishes at $r_A$). For
negative $\alpha$ we have $r_A>r_s>r_B$  being $r_A$ and $r_B$ given by the
same
equations.

More surprisingly, when the scalar
curvature is computed for the modified metric, it is found to be divergent at
$r_A$ and $r_B$. This shows the existence of an intrinsic singularity at least
in the most external radius. In addition, the signs of these
 divergences are different in such a way that in the limit $\alpha$ going to
zero (where the standard  Schwarzschild solution must be found) we see that
$r_A$ and $r_B$ go both to $r_s$.  In this limit the corresponding two
divergences
of the scalar curvature cancel with each other because of the different sign,
the
singularity disappear as it must be and the $A=B^{-1}$ relation is recovered.

\vskip 1.0cm

{\bf The modified Robertson-Walker metric:}

The most general isotropic and homogeneous metric can be written as:
\begin{equation}
d\tau ^2=dt ^2-a^2(t)(\frac{dr^2}{1-kr^2}+r^2d\theta^2+r^2sin^2\theta d\phi^2)
\end{equation}
where $k=-1,0,1$ corresponding to an open, flat or closed space and $a(t)$
is the scale parameter which
must be determined by the equations of motion. Any isotropic and homogeneous
distribution of  matter will be described by a energy momentum tensor that
takes  the form of a perfect fluid:
\begin{equation}
T^{\mu\nu}=(\rho + p)U^{\mu}U^{\nu}+pg^{\mu\nu}
\end{equation}
where $U^{\mu}$ is the velocity four vector, $\rho$ is the density, and $p$ is
the pressure. The energy conservation equation reads:
\begin{equation}
\frac{d(\rho a^3)}{da}=-3pa^2
\end{equation}
and finally, the state equation of matter is needed ($p=p(\rho)$) to
specify completely $a(t)$ from the equation of motion and the initial
conditions. In the case considered in this paper, the equations of motion
are:
\newpage \begin{eqnarray}
\frac{M_P^2}{16\pi}(-\frac{3}{a^2}(a'^2+k))+\frac{\alpha}{M_P^2}(-24\frac{a''^3}
{a^3}+12\frac{(a'^2+k)^3}{a^6}+
\end{eqnarray}
\begin{eqnarray}\nonumber
72(\frac{a'a''a'''}{a^3}-\frac{a'^2(a'^2+k)^2}
{a^6}))+\frac{\rho}{2}=0
\end{eqnarray}
which is the equation corresponding to the temporal coordinate. The three
spatial coordinates give rise to only one independent equation.

\begin{eqnarray}
\frac{M_P^2}{16\pi}(\frac{2a''}{a}+\frac{a'^2+k}{a^2})+\frac{\alpha}{M_P^2}(
12\frac{(a'^2+k)^3}{a^6} \\
\nonumber
-72(\frac{a''''a''+a'''^2}{3a^2}-\frac{a''(a'^2+k)(5a'^2+k)}
{3a^5}+\frac{a'^2(a'^2+k)^2}
{a^6}))+\frac{p}{2}=0\end{eqnarray}
Using  the energy conservation equation it is not difficult to see that
the derivative of the first equation above is in fact the addition of
the two equations. Therefore, we can take eq.18 as the only independent
equation.  However, this equation seems to be very difficult to solve. In
order to find some exact solutions we have tried to look for $a(t)$ functions
which also satisfy the relation
\begin{equation}
 a'^2+k=Ca^n
\end{equation}
The eq.18  forces $n$ to be equal 2 and:
\begin{equation}
\frac{M_P^2}{16\pi}C+\frac{4\alpha}{M_P^2}C^3=\frac{\rho}{6}
\end{equation}
Thus $\rho$ must be a constant and this fact, together with the conservation of
the energy equation imply the state equation $p=-\rho$. This means that the
kind
of solutions we are dealing with can describe matter only in the form
of a cosmological constant $\rho=\lambda M_P^2/8\pi$. In any case we can use
eq.21 to obtain the constant $C$ in terms of $\alpha$ and $\lambda$.
Independently of this value,eq.20 can be easily integrated with $n=2$ and
the following solutions are obtained:

a) In the case $C=0$ the eq.20  reads $a'^2=-k$ so for $k=0$ $a$ is constant
and
for $k=-1$, $a(t)=\pm t +t_0$. This case corresponds to the standard
Einstein equation with $\alpha=0$ in absence of matter.

b) For $C>0$ eq.20   reads $a'^2=Ca^2-k$. For $k=1, k=-1$ and $k=0$ the
solutions are respectively:
\begin{eqnarray}
a(t)=\frac{1}{\sqrt{C}}cosh(\sqrt{C}(t-t_0))     \\   \nonumber
a(t)=\frac{1}{\sqrt{C}}sinh(\sqrt{C}(t-t_0))  \\   \nonumber
a(t)=a(t_0)exp(\sqrt{C}(t-t_0))
\end{eqnarray}

c) For $C<0$ there is a solution only in the case $k=-1$ where
\begin{equation}
a(t)=\frac{1}{\sqrt{-C}}sin(\sqrt{-C}(t-t_0))
\end{equation}
We would like to stress that all these solutions satisfy exactly  eq.18 with
the Robertson-Walker metric provided that the $C$ is related to $\alpha$ and
$\lambda$ through eq.21 with $\rho=\lambda M_P^2/8\pi$. Therefore now we will
write $C$ in terms of the above mentioned constants. In particular we have the
important case of no matter and no cosmological term i.e. pure gravity where we
have three solutions: $C=0$, $C=\pm \sqrt{-M_P^4/64\pi\alpha}$. Thus, in this
case the $C$ different from zero solutions exist only for negative $\alpha$.

In the case of general $\lambda$ one must
consider the discriminant of the eq.21.
The sign of this discriminant  is in fact determined by $\alpha$ and
$\lambda$.  In any case, it is always possible to find an
analytical formula for the roots of eq.21 in terms of $\alpha$ and $\lambda$.

\vskip 1.0cm

{\bf Conclussions:}

At low-energies gravitation can be described by an effective lagrangian
which
is  an expansion in the number of derivatives over $M_P$. The first term of
this
expansion is the standard Einstein lagrangian but those with higher
derivatives have undetermined coefficients that finally could be computable
from
the underlying theory of gravitation in a similar way as the coefficients of
the
low-energy effective lagrangian for strong interactions could, at least in
principle, be obtained from QCD.

Independently of  which are the real values of these coefficients, the
consistence of the theory at low energies requires the $\alpha R^{\alpha\beta}
_{\;\;\;\;\gamma\delta}R^{\gamma\delta}_{\;\;\;\;\sigma\rho}
R^{\sigma\rho}_{\;\;\;\;\alpha\beta}$ term
considered in this work to be present in order to make it renormalizable in
the sense of the phenomenological lagrangians. Then it seems to be interesting
to
ask about the kind of effects that this term could produce and how it can
modify
the standard predictions of Einstein gravitation.

In the case of the gravitational field produced by a static and spherical
body, we obtain that the modifications to the large distance field are
rather weak and completely unobservable. However, in the region close to
the Schwarzschild radius, the qualitative behavior of the field changes
dramatically since the event horizon is transformed in an extended and
spherical singularity. In fact, as was already discovered in [11], this is a
very generic effect and it will appear virtually in any modification of the
field equations and not only in the one considered in this work. We think
that, at least to our knowledge, the physical consequences of this fact have
not
been yet  properly studied in the literature.

In the case of the cosmological solution we have shown that inflationary
solutions exist for the modified equations of motion which do not require the
somewhat artificial introduction of the inflaton field (this interesting
phenomenon was also found in [12] for the case of an extra $R^2$ term in the
gravitational lagrangian). Of course, in the real case one would need to
include the effect of higher order terms in  the equation of motion but
our simple solutions show that pure gravitation could produce inflation and
this could be relevant in  the very early history of the universe [13].
\vskip 1.0cm

{\bf Note added}

When this work was completed, we noticed reference [14] where the authors study
the effect of the six derivative terms on the Hawking temperature of black
holes.
\vskip 1.0cm
 {\bf Aknowledgments}

This work has been partially supported by the Ministerio de Educaci\'on y
Ciencia (Spain)(CICYT AEN90-0034).

{\newpage}


\thebibliography{references}


\bibitem{1} E. Alvarez {\em Rev. of Mod. Phys.} {\bf 61}, 561 (1989)

\bibitem{2} {\it Superstring theory }, M.B. Green, J.H. Schwarz and E. Witten.
Cambridge University Press (1987)

\bibitem{3}  S. Weinberg, {\em Physica} {\bf 96A} 327 (1979)

\bibitem{4}  J. Gasser and H. Leutwyler, {\em Ann. of Phys.} {\bf 158} 142
 (1984) , {\em Nucl. Phys.} {\bf
B250} 465 and 517 (1985)

\bibitem{5}  A. Dobado and M.J. Herrero, {\em Phys. Lett.} {\bf B228}
495  (1989)   and {\bf B233} 505 (1989)  \\
 J. Donoghue and C. Ramirez, {\em Phys. Lett.} {\bf B234} 361 (1990)

\bibitem{6}  N. Nakanishi and I. Ojima, {\em Phys. Rev. Lett.} {\bf 43}
91  (1979)

\bibitem{7}  G. 't Hooft and M. Veltman, {\em Ann. Inst. H. Poincar\'e} {\bf
20} 69  (1974)


\bibitem{8}  M.H. Goroff and A. Sagnotti, {\em Nucl. Phys.} {\bf
B266} 709
 (1986)

\bibitem{9}  D.M. Capper, J.J. Dulwich and M. Ram\'on Medrano, {\em Nucl.
Phys.}
{\bf B254} 737
 (1985)

\bibitem{10} {\it Gravitation and Cosmology}, S. Weinberg, John Wiley   \& Sons
(1972)

\bibitem{11}  W. Israel, {\em Phys. Rev.} {\bf 164} 1776 (1967)
 200

\bibitem {12} A.A. Starobinsky, {\em Phys.
Lett.}
 {\bf 91B} 99 (1980)

\bibitem {13} A. Dobado and A. L\'opez, in preparation.


\bibitem{14} M. Lu and M.B. Wise, {\em Phys. Rev. } {\bf D47}  3095 (1993)
\end{document}